\documentclass[aps,prl,a4paper]{revtex4}

\usepackage{graphicx}

\usepackage{fancyhdr}
\usepackage{pslatex}

\begin{document}
\title {Effects of an additional dimension in the Young experiment}

\author{Allan Kardec Barros }

\affiliation{Universidade Federal do Maranhao}

\begin{abstract}
The results of the Young experiment can be analyzed either by classical or Quantum Physics. The later one though leads to a more complete interpretation, based on two different patterns that appear when one works either with single or double slits. Here we show that the two patterns can be derived from a single principle, in the context of General Relativity, if one assumes an additional spatial dimension to the four known today. The found equations yield the same results as those in Quantum Mechanics.
\end{abstract}

\vskip -1.35cm

\maketitle

\thispagestyle{fancy}

\setcounter{page}{1}

\bigskip

\section{I. INTRODUCTION}

The wave-particle duality defies intuition, as it appeared in many studies\cite{Walborn,Kim,Tonomura,Gerlich}. Let us take the classical Young double-slit experiment. It basically consists of making a light source to focus on a thin plate with two parallel slits, with a screen behind them (as in Fig 1). It is well known that the experiment works in the following fashion: 1) If only one slit is open (or, one shut), after a large number of particles have reached the screen, one should find a non-periodic pattern. This is regarded as "particle behaviour"\cite{Walborn,Kim}.  If the two slits are open. One finds a periodic pattern. This case is regarded as "wave behaviour". Even if we send one photon at a time, the phenomenon repeats. Still, one is forced to wait until a large number has been sent, to characterize the well-known periodicity, as shown in Tonomura experiment\cite{Tonomura}.

The theoretical prediction of the pattern appearing at the screen are historically well described and precise, either using classical or Quantum Physics. In fact, in the case of one slit, the pattern can be described by the Fraunhofer diffraction equation. While in the two slits case, one can use a squared cosine function modulated by a sinc function. It is interesting though that the pattern at the screen is formed by small dots, which thus explains the particle aspect. It is even more intriguing the results of a study where it was observed the averaged trajectories of single photons\cite{Kocsis}, suggesting a particle trajectory.

	It is also important to notice the large growth of interest on extra dimensions recently.   A number of issues appeared suggesting the increase of the number of dimensions, which yielded the appearance of String theory, M-theory\cite{Becker,Leon} and many other related theories, generally called ÒKaluza-Klein theoriesÓ. An interesting review was carried out by Maartens and Koyama\cite{Maartens}. However, one of the important points of those theories remains critical, which is the lack of real world evidences to ground them, a problem faced here. Moreover, there are different works in the literature regarding the physical interpretation of the 5th dimension. Indeed, it is not always length-like, at it could be also mass\cite{Wesson1} with various consequences, with one which is particularly important for this work is that studied by Wesson\cite{Wesson1}, which  describes waves in vacuum.
	
The fact is that, although an old one, the Young experiment is still used today to demonstrate the claim of duality. This would work in a complementary way: either we would find a particle or a wave behaviour Ð not both at the same time. Indeed, a sharp transition would split the two behaviours. In order to explain those phenomena using a single principle, we propose to increase the number of dimensions basically in the same way of that of Kaluza and Klein\cite{Overduin,Wesson}. In this regard, Kaluza first suggested an additional dimension to include the electromagnetic forces\cite{Overduin,Wesson}. Then, answering to the question of why the fifth dimension was not observed, Klein proposed that this dimension would have a circular topology with its radius proportional to the Planck length\cite{Overduin,Wesson}. To our knowledge, the large majority of the consequent theories are grounded on the later assumption. In this work, we drop that. The additional dimension should also have a circular topology, but with a radius proportional to the particle wavelength, or, to be more appropriate, we take the De Brogile's relation and use the term "wavelength". Here we follow the strategy in the literature of connecting the 5th dimension to properties of the particle, for example, to mass\cite{Wesson}. With this, the trajectories of particles in the 3D space are rotated by an angle proportional to the shifts in the circular dimension. The two behaviours can then be predicted simply by estimating the trajectories of individual particles.

\section{II. METHODS}

Let us define a five dimensional coordinate vector, $x$, and the corresponding spacetime metric tensor,
\begin{equation}
g_{AB}= 
\left ( 
\begin{array}{cc}
g_{\alpha\beta}+\kappa^2\phi A_{\alpha} A_{\beta} & \kappa\phi A_{\alpha}\\
 \kappa\phi A_{\beta}& \phi
\end{array}
\right ).
\label{eqn:metric}
\end{equation}
where $g_{\alpha\beta}$ is the fourth dimensional metric tensor,  $A_{\alpha}$ is the electromagnetic potential, $\kappa$ is a scaling parameter and $\phi$  is a negative scalar field. We used Greek letters to characterize the four dimensional spacetime. The four-dimensional metric signature is taken to be $(-+++)$, and we work in units such that $c = 1$.  We assume therefore that the fifth coordinate has a circular topology ($S^1$), and is periodic in $ y$, where the radius changes according to the relation $\lambda=h/p$, where $p$ is the linear momentum of the particle and $h$ is the Planck constant. Thus, the fields, $f(y)$, become periodic and they can be rewritten in Fourier terms as follows,
\begin{equation}
f^{A}(y)=\sum_{i=-\infty}^{\infty}\hat f^{A}(i)e^{({ji\pi y}/{\lambda})},
\label{eqn:fourier}
\end{equation}
where $\hat f^{A}(i)$ are Fourier coefficients. 

When there are no external masses nor charges, $g_{\alpha\beta}=\eta_{\alpha\beta}$  and $A_{\alpha}= 0$, where $\eta_{\alpha\beta}$ is the Minkowski metric. Remembering that $\eta_{\alpha\beta}$ is part of a Lorentz group, i.e., $\eta_{\alpha\beta}=\Psi(y)_\alpha^{\rho}\Psi(y)_\beta^{\sigma}\hat\eta_{\rho\sigma}$, the theorem below shows that $\Psi(y)_\alpha^{\rho}$ should be a rotation matrix.

{\bf Theorem}
When one of the coordinates, e.g., $y$, have a circular topology, and the metric is that of Minkowski, $\eta_{\alpha\beta}$,  then there is a rotation matrix $\Psi(y)_\delta^{\alpha '}$, which obeys $\eta_{\alpha\beta}=\Psi(y)_\alpha^{\rho}\Psi(y)_\beta^{\sigma}\hat\eta_{\rho\sigma}$, and whose elements are periodic.

{\bf Proof.} We shall develop our reasoning in parts.

\begin{enumerate}

\item {\it Orthogonality restriction}. Let us first remember that $g_{\alpha}^\beta(i,y)$ should be orthogonal, as its non-diagonal elements are null. Thus, we have to rewrite $ g_{\alpha}^\beta(i,y)$ in terms of the Fourier series (\ref{eqn:fourier}), taking this restriction into account. Luckily, each spectral line of the Fourier series are mutually orthogonal.
Without loss of generality, let us analyze a simple case  when $ g_{\alpha}^\beta(i,y)$ is a $2 \times 2$ matrix. Moreover, let us take only one spectral line of the Fourier expansion. Using  (\ref{eqn:fourier}), $ g_{\alpha}^\beta(i,y)$ can be rewritten as,
\begin{eqnarray}
 g_{\alpha}^\beta(i,y)=  
\left ( 
\begin{array}{cc}
a_{1}^1 \cos(i\pi y/\lambda)  + b_{1}^1\sin(i\pi y/\lambda) & b_{2}^1\sin(i\pi y/\lambda) + a_{2}^1\cos(i\pi y/\lambda)\\
a_{1}^2 \cos(i\pi y/\lambda)  + b_{1}^2\sin(i\pi y/\lambda) & b_{2}^2\sin(i\pi y/\lambda) +  a_{2}^2\cos(i\pi y/\lambda).
\end{array}
\right )
\label{eqn:cosine}
\end{eqnarray}

\item Any neighboring elements of (\ref{eqn:cosine}) should be mutually orthogonal.  By making 
\begin{eqnarray}
a_{1}^1=b_{2}^1,
a_{2}^1=-b_{1}^1,
a_{1}^2=b_{2}^2,
a_{2}^2=-b_{1}^2.
 \label{eqn:elements}
 \end{eqnarray}
we force mutual orthogonality of any two column elements in (\ref{eqn:cosine}).

\item However, we still have to assure orthogonality between lines. Inserting (\ref{eqn:elements}) in  (\ref{eqn:cosine}) we have,

\begin{eqnarray}
 \label{eqn:cosinemult}
 \hat g_{\alpha}^\beta(i,y)=  
\left ( 
\begin{array}{cc}
a_{1}^1 & a_{2}^1\\
a_{1}^2 & a_{2}^2
\end{array}
\right )
\left ( 
\begin{array}{cc}
\cos(i\pi y/\lambda)  & \sin(i\pi y/\lambda) \\
-\sin(i\pi y/\lambda) & \cos(i\pi y/\lambda) 
\end{array}
\right )=\Omega_{\alpha}^\rho(i,y)\Psi_{\rho}^\beta(i,y).
 \end{eqnarray}

Obviously, $\Psi_{\alpha}^\beta(i,y)$  is a rotation matrix. By making $\Omega_{\alpha}^\beta(i,y)=\Psi(i,y)^{\alpha}_\beta(i,y)$, $\forall (i,y)$, leads to 
\begin{equation}
\Psi(i,y)_{\alpha}^\rho\Psi(i,y)_{\rho}^\beta=\delta_{\alpha}^\beta,
\label{eqn:kroneker}
\end{equation}
which is an orthonormal matrix.

\item We can have only one spectral element in the Fourier series, as in (\ref{eqn:fourier}).  Any additional element will decrease the unity value present in $\delta_{\alpha}^\beta$.

\item The simplest way to find (\ref{eqn:kroneker}) is by making either $y=0$ or $i=0$. Notice that both are included in (\ref{eqn:cosinemult}). Thus, we have,
\begin{equation}
 g_{\alpha\beta}=\eta_{\alpha\beta}=\Psi(i,y)_\alpha^{\rho}\Psi(i,y)_\beta^{\sigma}\hat\eta_{\rho\sigma}
\label{eqn:neta}
\end{equation}

\item Notice that (\ref{eqn:neta}) is not fulfilled if we include the time coordinate, which is negative in the Minkowski metric. Thus, we shall deal only with the spatial elements. Indeed, if we add one more spatial coordinate, (\ref{eqn:neta}) is still valid and it is not difficult to generalize these results to any $n \times n$ matrix. $\diamondsuit$

\end{enumerate}

The result above means that we should have,
\begin{equation}
x^{\alpha}=\Psi(i,y)_\rho^{\alpha } x^\rho .
\label{eqn:rot}
\end{equation}

{\bf Remarks}
\begin{enumerate}

\item The immediate consequence of this result is that, although the metric is preserved, the rotation matrix will act on the spatial coordinates. If we fix $y$, we will have a fourth dimensional spacetime. In this regard, the final effect of the rotation matrix in (\ref{eqn:rot}) is to change the direction of travel of a given particle in an $x$ plane according to the shifts of $y$,  i.e., for $\Delta y \neq 0$ we have $\Psi(i,y)_\rho^{\alpha }\neq \delta_\rho^{\alpha }$. 

\item By looking at (\ref{eqn:rot}), one can see that any angle in the argument of the sinusoidal functions should be embedded in the respective plane. This means that any shift in the extra coordinate, say, $\Delta y$, can be measured in the plane the rotation happens.

\end{enumerate}

\subsection{III. SINGLE AND DOUBLE SLITS CASES}

Let us take the general case where a particle departs from a source, crosses an infinitely thin slit and hits a screen. 
Thus, the particle can take two different paths with their corresponding directions. One from the source to the slit (which we call {\it first part}) and another from the slit to the screen ({\it second part}).  Thus, each direction changes by, say, $\cos(\pi \Delta y_i/\lambda)$, where $\Delta y_i$ is the $i-th$ shift in the $y$ coordinate.
Let us assume that the particle can take $N$ different directions.
We take the probability, $\psi_i$, of the particle to reach a given point and relate it to the shifts in the trajectory,  $\psi_i=(1/N)\cos(\pi \Delta y_i/\lambda)$. 

Therefore, we should have a product of probabilities of the particle taking one of the paths in the first part, by the probability of the particle taking one of the paths in the second part. Additionally, the probabilities obey the identity $\cos(\varpi)=\cos(-\varpi)$.  For each part, we will only use straight paths, which means that for $N$ slits, we will have $N$ different paths for the second part, which also defines the number of possible paths in the first part.  Thus, the matrix $\Upsilon_{i}^j$ below shows the  possible combinations,
\begin{eqnarray}
\Upsilon_{i}^j=  
\left ( 
\begin{array}{ccccc}
\psi_1\psi_1 & \psi_1\psi_2 &  \cdots  & \psi_1\psi_N \\  
\psi_2\psi_1 & \psi_2\psi_2 &  \cdots  & \psi_2\psi_N\\ 
\vdots & \vdots & \ddots &  \vdots  \\ 
\psi_N\psi_1 & \psi_N\psi_2 &  \cdots  & \psi_N\psi_N\\ 
\end{array}
\right ).
\label{eqn:particao}
\end{eqnarray}

Finally, the probability at a given point will be the sum of the probabilities of the particle to cross each slit, which is that of all possible combinations. Thus, we take the number of particles to be a weighted sum of the elements of $\Upsilon_{i}^j$ in  (\ref{eqn:particao}),  given by,
\begin{eqnarray}
\Xi_{N} ={\bf 1}^i{\bf 1}_j\Upsilon_i^j=\frac{1}{N^2}\left [{\bf 1}^i\cos(\pi \Delta y_{i}/\lambda)\right ]^2,
\label{eqn:proposta1}
\end{eqnarray}
where ${\bf 1}_j $ is a vector of ones and the maximum number of particles, $\Xi_{max}$, was set to $\Xi_{max}=1$, for convenience.

We now face the two cases: single and double slits. However, actual slits have a width, which makes the problem a bit more complex. Thus, we present firstly an example to highlight our reasoning, 
and then generalize for actual single and double slits.

\subsubsection{III. 2. An example}

Let us take as an example matrix $\Upsilon_{i}^j$ with two terms, $\psi_1$ and $\psi_2$, in (\ref{eqn:particao}). Moreover, let us assume that we are dealing with two infinitely thin slits, as shown in Fig.~1. The task is to count the number of particles that reach point $q$.
As stated above, we make a simple assumption which is that the paths are straight lines and the only possible paths in the $y$ coordinate are those determined by the second part to reach point $q$.
In Fig.~1, we show that for two slits we can have two different angles:  $\theta_1$ and $\theta_2$. We show three different trajectories: two of them are straight lines, one given by $\theta_1$ and the other by $\theta_2$, and the third one is a composition of $\theta_1$ and $\theta_2$.
We assume the angles $\theta_i$ to be very small, as in the literature~\cite{Caruso}. 
For a non-zero angle $\hat\theta$, we shall have a non-zero shift in the $y$ coordinate.
We shall associate also the shifts in $y$ coordinate to the observed ones on the  $x$-plane. For small shifts, $\xi$, we can use the approximation $\xi =\alpha\Delta y_j+\beta$. Moreover, this relation  should obey $\cos(\pi \Delta y_i/\lambda)=\cos(\pi\xi/\lambda)$, whose solution is $\xi =\Delta y_j\pm n 2\pi$, where $n$ is an integer.
If we approximate $\xi$ to the arc length, then, $\xi\approx a\theta =\Delta y_j$, where we used $n=0$. 
 Then, defining $\theta_2-\theta_1=2\hat\theta$, we can use  (\ref{eqn:particao}) and find,
\begin{eqnarray}
\Upsilon=  
\left ( 
\begin{array}{cc}
\cos \left( {\pi a\theta_1}/2\lambda \right) \cos \left( {\pi a\theta_1}/2\lambda \right) 
& \cos \left( {\pi a\theta_1}/2\lambda \right) \cos \left( {\pi a\theta_2}/2\lambda \right)   \\ 
 \cos \left( {\pi a\theta_2}/2\lambda \right) \cos \left( {\pi a\theta_1}/2\lambda \right) 
& \cos \left( {\pi a\theta_2}/2\lambda \right) \cos \left( {\pi a\theta_2}/2\lambda \right)   \\  
\end{array}
\right ).
\label{eqn:particao2}
\end{eqnarray}

From this and substituting (\ref{eqn:particao2}) in (\ref{eqn:proposta1}),  we observe that the contribution, $\Xi_{2} $, can be written as,
\begin{eqnarray}
\Xi_{2} =\frac{1}{4}
\left [\cos \left( \frac{\pi a\theta_1}{2\lambda} \right)+
\cos\left(  \frac{\pi a\theta_2}{2\lambda} \right)\right ]^2
\label{eqn:energy}
\end{eqnarray}

If we make $\theta_1=\omega-\hat\theta$ and $\theta_2=\omega+\hat\theta$, and substitute in
 (\ref{eqn:energy}) we  have,
\begin{eqnarray}
\Xi_{2}\propto
\cos^2\left(  \frac{\pi a\hat\theta}{\lambda}\right),
\label{eqn:energyfinal}
\end{eqnarray}
which is equivalent to the result found in the literature\cite{Caruso}.

\subsubsection{III. 3. The single slit case}

With this result on our hands, we can find the equation to the case of one actual slit. Let us remember again that every actual slit has a finite width $d$, although ideally we modeled it above as infinitely thin. With this in mind, we understand the one slit case as the result of the contribution of a large number of ideal slits. However, with the distance between any two of them bounded to the width of that actual single slit. Thus, we can model the actual single slit as a sum of $N$ ideal slits, evenly spaced with width $d/N$. 
We use here the same reasoning to arrive at (\ref{eqn:energyfinal}), i.e., given the width of $i$-th slit, $i d/N$, we assume that we can approximate each difference in the trajectory by its corresponding arc length.
Using the right side of (\ref{eqn:proposta1}) and following the steps as in the example above,  we can write the following equation, 
\begin{eqnarray}
\Xi_{1slit}\propto\frac{1}{N^2}
\left [\sum_{i=0}^N \cos\left(i\frac{d\pi}{\lambda N}\hat\theta\right)\right ]^2
\label{eqn:energytotals1}
\end{eqnarray}

For this, let us use of a known relation on sum of cosines which is,
\begin{eqnarray}
\sum_{i=0}^N \cos\left(i\frac{1}{N}\hat\theta\right)=
\frac{\cos(\frac{N}{2}\frac{1}{N}\hat\theta)\sin(\frac{N+1}{2}\frac{1}{N}\hat\theta)}{\sin(\frac{1}{2}\frac{1}{N}\hat\theta)}
\approx
\frac{N\sin(\hat\theta)}{\hat\theta}
=
N{\rm sinc}(\hat\theta),
\label{eqn:sumcos}
\end{eqnarray}
where we assumed $N>>1$ and a very small $\hat\theta$.
Thus, using (\ref{eqn:energyfinal}), (\ref{eqn:energytotals1}) and (\ref{eqn:sumcos}), the  number of particles will be, 
\begin{eqnarray}
\Xi_{1slit}\propto
{\rm sinc}^2\left(\frac{\pi d\hat\theta}{\lambda}\right),
\label{eqn:sumcos2}
\end{eqnarray}
where $d$ is the width of a single slit. 

\subsubsection{III. 4. The double slit case}

	In the double slit case, we use the same reasoning as that used for the single slit. The only difference now is how we model it. Basically, we take a large slit and block part of it. The reasoning is the same as used to obtain (\ref{eqn:sumcos2}). We carry out three operations: 1) we generate one slit with width $a+d$, with $a>>d$, and; 2) build another one with width $a-d$; 3) Finally, we remove the latter from the first, which is equivalent to adding a block to the first one. Using (\ref{eqn:sumcos}), we can model this architecture as,
 \begin{eqnarray}
\frac{\sin((a+d)\vartheta)}{(a+d)\vartheta}-\frac{\sin((a-d)\vartheta)}{(a-d)\vartheta}\approx
\frac{2\sin(a\vartheta)\cos(d\vartheta)}{a\vartheta}=
2\cos(d\vartheta){\rm sinc}(a\vartheta).
\label{eqn:sumcos3}
\end{eqnarray}	

Accordingly, by using the strategy carried out before, we have that the resulting number of particles will be given by,
 \begin{eqnarray}
\Xi_{2slits}\propto
\cos^2\left(\frac{\pi a\hat\theta}{\lambda}\right){\rm sinc}^2\left(\frac{\pi d\hat\theta}{\lambda}\right),
\label{eqn:energytotal2}
\end{eqnarray}
which is again the same result as in the literature, where $a$ is the distance between the slits and $d$ is their width.

\section{III. DISCUSSIONS}

In this work, we saw that one can build the description of a trajectory behavior in the Young experiment based on geometry. We propose a theorem which shows that when the metric is that of Minkowski, where one spatial coordinate has a circular dimension, rotations occur in the Euclidian space, which is a consequence of (\ref{eqn:rot}). With this result, we could find an equation for the number of particles arriving from two infinitely thin slits, described by (\ref{eqn:energyfinal}). This leaded to the equations for either one or two actual slits.
This is based on the fact that one can build either a periodic or a sinc pattern only by using sum of cosines. Indeed, we found equations (\ref{eqn:sumcos2}) and (\ref{eqn:energytotal2}) which shows that the number of particles $\Xi_{1slit}$, for one actual slit, and $\Xi_{2slits}$, for two actual slits,  are just the same as those found in Quantum Mechanics, with the squared number of particles replaced by probabilities. 

Another consequence of the rotation in (\ref{eqn:rot}) is asymmetry in time. Indeed, the rotation matrix has  the circular coordinate as argument, which is clearly a non-linear operation. This means that any shift in the extra coordinate has, as consequence, a shift in the non-cylindrical ones. Those shifts are not, however, symmetrical, which means that we will not have any longer a symmetry in time (which occurs without the extra coordinate).
Obviously, if we use the definition of entropy as function of volume, a shift in the extra coordinate will always cause an increase in entropy. This fact might lead to a proof for the second law of thermodynamics and to a possible explanation for the arrow of time.

An interesting aspect is when we predict the outcomes of a sum of $n$ cosines. For this, we carried out some simulations where we added different numbers of cosines. In Fig.~3, we see the error between the sinc function and a sum of up to 20 cosine functions. We can also see that as few as 10 terms could clearly build a sinc function. However, even fewer terms could be easily confused with an interference pattern, as we can see in upper row ($n=3$ and $n=10$). The pattern is also altered if we add a small nuisance to the sum of cosines, as seen in Fig.~3 (lower row). The interpretation we give to this noise is equivalent to obstacles in the  particleÕs trajectory\cite{Walborn,Kim}, caused, for example, by the refraction on the material it is crossing. 
Additionally, we can easily predict the three-slits outcome using the reasoning above\cite{Sinha}. 

One might ask what happens if we have particles with different moments. We have seen that the shift in the trajectory changes according to the wavelength. We shall expect therefore a sinc-like distribution even with a small number of particles, as the sum of cosines tends rapidly to a sinc function. Another matter regards the use of the particle properties in the metric. As we said, we are not bringing any new idea in this regard, as it was already used in the literature\cite{Wesson}. In fact, a large discussion is carried out by Wesson\cite{Wesson} to the limits of using properties of the particle in the metric, the conclusion was that the Weak Equivalence Principle would not be any longer valid. This is not the focus of this work, however, one can reach simple conclusions. One of those is that for very heavy objects (i.e., of cosmological mass) the fields in (\ref{eqn:fourier}) will tend to a constant, yielding therefore the cylinder condition. However, it remains unclear today how the acceleration would work in periodic fields.

Let us also be clear about a possible confusion: our approach is not that of Quantum Mechanics and therefore we do not rely on wave functions. Rather, our proposal is based in a five dimensional metric, with the consequences that follows due to the propagation of a particle: a defined trajectory. One would argue that this is a problem, as there might be no evidences for those "classical" trajectories. However, the work by Kocsis et al \cite{Kocsis} with single photons showed evidences  that photons have well defined trajectories in the Young experiment.

The proposed framework, although applied here to the Young experiment, can be used in many other problems.  One of them is {\it entanglement}, a task that we do not regard as theoretically difficult in our framework. A quite challenging  experiment would be to measure the time that entangled particles would take from one point to another, which is still controversial\cite{Salart,Zeilinger}. 
Our work states clearly that the particles would take well defined trajectories and therefore one can predict straightforwardly that time interval.

Moreover, there is a result which appeared in the literature, showing a quite different pattern, which is that of a continuity, rather than a sharp transition from {\it particle} to {\it wave} \cite{Peruzzo,Kaiser}. 
Our work also appeals to that matter, if we simulate, rather than with two slits with equal widths, two ones with distinct widths.
Additionally, we believe that, by using similar designs such as that of the {\it delayed choice quantum eraser}\cite{Walborn,Kim}, and by precisely adjusting the obstacles, we can predict the outcomes as a {\it continuity} between {\it particle} and {\it wave}.


\section{ACKNOWLEDGMENTS}

I am deeply indebted to Professor Luiz N. Oliveira for his highly qualitative discussions.

\newpage

\begin{figure}[!htb1]
\begin{center}
\includegraphics*[angle=0,  width=16cm]{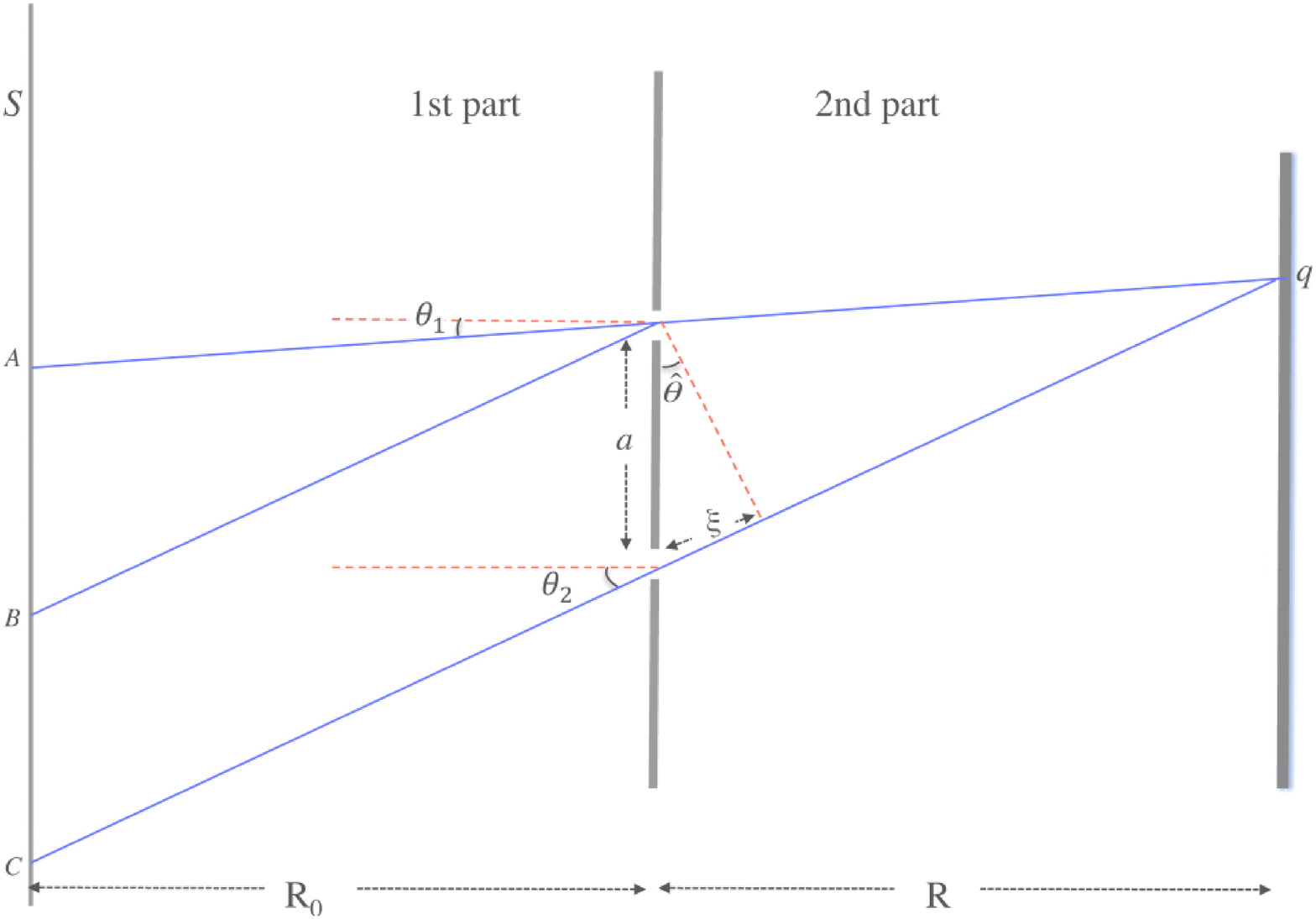}
\end{center}
\caption{{An example of Young experiment for two ideal slits. A particle departs from a source $S$ and crosses  one of two different slits, reaching a point $q$. Each path can be split in two others. One from the source to the slits (1st part), and another from the slits to the screen (2nd part). 
The particle takes paths  making either an angle $\theta_1$ or an angle $\theta_2$ to the direction perpendicular to the screen. Here we highlight three possible trajectories: two with the same angle (departing from A and C) and another one where the angle changes after the slit (departing from B). 
The  distance between the slits is given by $a$, while $\xi$ is the  difference between the two trajectories, linked to angle $\hat\theta$, and $\{R,R_0\}>>a$.
The classical result~\cite{Caruso} shows that the intensity at the screen is proportional to ${\rm cos}^2(\pi a \hat\theta/ \lambda)$.
}}
\end{figure}

\begin{figure}[!htb1]
\begin{center}
\includegraphics*[angle=0, width=14cm]{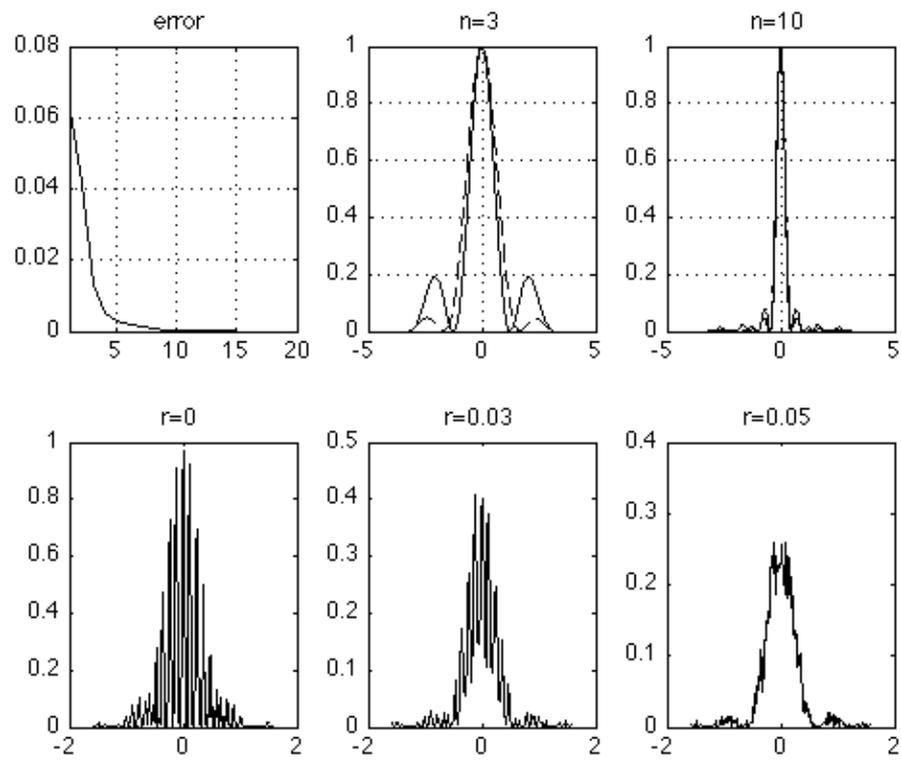}
\end{center}
\caption{Examples of sum of cosines in different situations. Upper row: (left) Approximation error between a sinc function and sum of cosines for 20 terms and illustration for two different number of terms, as function of the angle (in rad.), and (right) two examples of sinc functions for $n=3$ and $n=10$; Lower row: Simulation of a two slit experiment when one disturbs the trajectory by a noise at different standard deviations: 0.0, 0.03 and 0.05.}
\end{figure}

\end{document}